\documentclass[prl,a4paper,superscriptaddress,twocolumn,amsmath,amssymb,floatfix,amstex]{revtex4-1}%
\newcommand{\ket}[1]{| #1\rangle}                       %

\newcommand{\ktimes}{\rangle\! \langle}
\newcommand{\op}[2]{|#1\ktimes #2|}
\def\mpool{M_\textrm{pool}}
\def\img{\textrm{Im} G}
\def\nev{\lambda}

\usepackage{graphicx,graphics,rotating}	
\usepackage{dcolumn}
\usepackage{bm}
\usepackage{hyperref}
\usepackage{color}
 \usepackage{soul}   								

\begin{document}
\title{Scaling theory of the Anderson transition in random graphs:\\ ergodicity and universality}

\author{I.~Garc\'ia-Mata}
\affiliation{Instituto de Investigaciones F\'isicas de Mar del Plata
(IFIMAR), CONICET--UNMdP,
Funes 3350, B7602AYL
Mar del Plata, Argentina.}
\affiliation{Consejo Nacional de Investigaciones Cient\'ificas y
Tecnol\'ogicas (CONICET), Argentina}
\author{O.~Giraud}
\affiliation{LPTMS, CNRS, Univ.~Paris-Sud, Universit\'e Paris-Saclay, 91405 Orsay, France}
\author{B.~Georgeot}
\affiliation{%
Laboratoire de Physique Th\'eorique, IRSAMC, Universit\'e de Toulouse, CNRS, UPS, France
}
\author{J.~Martin}
\author{R.~Dubertrand}
\affiliation{Institut de Physique Nucl\'eaire, Atomique et de
Spectroscopie, CESAM, Universit\'e de Li\`ege, B\^at.\ B15, B - 4000
Li\`ege, Belgium}
\author{G.~Lemari\'e}
\email[Corresponding author: ]{lemarie@irsamc.ups-tlse.fr}
\affiliation{%
Laboratoire de Physique Th\'eorique, IRSAMC, Universit\'e de Toulouse, CNRS, UPS, France
}

\date{September 9, 2016}%
\begin{abstract}
We study the Anderson transition on a generic model of random graphs with a tunable branching parameter $1<K\le 2$, through large scale numerical simulations and finite-size scaling analysis. We find that a single transition separates a localized phase from an unusual delocalized phase which is ergodic at large scales but strongly non-ergodic at smaller scales. In the critical regime, multifractal wavefunctions are located on few branches of the graph. Different scaling laws apply on both sides of the transition: a scaling with the linear size of the system on the localized side, and an unusual volumic scaling on the delocalized side. The critical scalings and exponents are independent of the branching parameter, which strongly supports the universality of our results.  
\end{abstract}

\maketitle

Ergodicity properties of quantum states are crucial to assess transport properties and thermalization processes. They are at the heart of the eigenstate thermalization hypothesis which has attracted enormous attention lately~\cite{Jensen85,*Deutsch91,*Srednicki1994,*Calabrese2006,*Rigol2008}. A paramount 
example of non-ergodicity is Anderson localization where the interplay between disorder and interference leads to exponentially localized states \cite{anderson1958absence}. 
In 3D,  a critical value of disorder separates a localized from an ergodic delocalized phase. At the critical point eigenfunctions are multifractal, another non trivial example of non-ergodicity \cite{abrahams201050, evers2008anderson}. 
Recently, those questions have been particularly highlighted in the problem of many-body localization \cite{jacquod1997emergence, basko2006metal,pal2010many,nandkishore2015many,altman2015universal}. Because Fock space has locally a tree-like structure, the problem of Anderson localization 
on different types of graphs \cite{abou1973selfconsistent, efetov1985anderson, zirnbauer1986localization, *zirnbauer1986anderson, castellani1986upper,mirlin1991universality, *fyodorov1991localization, *fyodorov1992novel, mirlin1994distribution} has attracted a renewed activity \cite{monthus2008anderson,monthus2011anderson,biroli2012difference,deluca2014anderson,kravtsov2015random,altshuler2016non,tikhonov2016anderson,facoetti2016non,monthus2016localization, tikhonov2016fractality}. In particular, the existence of a delocalized phase with non-ergodic (multifractal) eigenfunctions lying on an algebraically vanishing fraction of the system sites is debated \cite{deluca2014anderson,kravtsov2015random, altshuler2016non, facoetti2016non, tikhonov2016fractality}.

The problem of non-ergodicity also arises in another context corresponding to glassy physics \cite{mezard1990spin}. 
For directed polymers on the Bethe lattice \cite{derrida1988polymers}, a glass transition leads to a phase where a few branches are explored among the exponential number available. As there is a mapping to directed polymer models in the Anderson-localized phase \cite{abou1973selfconsistent, miller1994weak, somoza2007universal, monthus2009statistics}, it has been recently proposed that this type of non-ergodicity (where the volume occupied by the states scales logarithmically with system volume) could also be relevant in the delocalized phase \cite{biroli2012difference}. Note however that it has been envisioned  that this picture could be valid only up to a finite but very large length scale \cite{biroliprivate}.

In this letter, we study the Anderson transition (AT) in a family of random graphs \cite{ZhuXiong00, ZhuXiong01, Gir05}, where a tunable parameter $p$ allows us to interpolate continuously between the 1D Anderson model and the random regular graph model of infinite dimensionality. 
Our main tool is the single parameter scaling theory of localization \cite{abrahams1979scaling}. 
It has been used as a crucial tool to interpret the numerical simulations of Anderson localization in finite dimensions \cite{pichard1981finite,mackinnon1981one,abrahams201050,Slevin99} and to achieve the first experimental measurement of the critical exponent of the AT in 3D \cite{chabe2008experimental}. In our case, the infinite dimension of the graphs leads to highly non-trivial finite-size scaling properties: unusually, we find different scaling laws on each side of the transition.  
Our detailed analysis of extensive numerical simulations leads to the following scenario. A single AT separates a localized phase from an ergodic delocalized phase. However a characteristic non-ergodicity volume (NEV) $\Lambda$ emerges in the latter phase. For scales below $\Lambda$, states are non-ergodic in the sense that they take significant values only on few branches, on which they additionally display multifractal fluctuations. For scales above $\Lambda$, this structure repeats itself and leads to large scale ergodicity. At the threshold, $\Lambda$ diverges, and the behavior below $\Lambda$ extends to the whole system. The critical behaviors do not depend on the graph parameter $p$, which strongly supports the universality of this scenario.

In order to describe the localization properties, we use two complementary approaches. First we derive recursive equations for the local Green function using a mapping to a tree \cite{abou1973selfconsistent}, which we solve using the pool method from glassy physics \cite{miller1994weak, monthus2008anderson}, and analyse the critical behavior by finite-size scaling. Second, we perform exact diagonalization of very large system sizes up to $N\approx 2 \times 10^6$, and we extract the scaling properties of eigenfunction moments. We use the box-counting method in this new context of graphs of infinite dimensionality to perform a local analysis and to extract the NEV $\Lambda$ unambiguously.

\emph{Random graph model.---} 
We consider a 1D lattice of $N$ sites with periodic boundary conditions. Each site is connected to its nearest neighbors  and  $\lfloor pN \rfloor$ shortcut links are added ($\lfloor . \rfloor$ is the integer part). These shortcuts give an average distance between pairs of sites that increases logarithmically with $N$, so that the graph has an infinite dimensionality (see Supplemental Material and \cite{Str00}). 
The system is described by an $N$-dimensional Hamiltonian $H=\sum_{i=1}^N\varepsilon_i\op{i}{i}+\sum_{\langle i,j\rangle} \op{i}{j}+\sum_{k=1}^{\lfloor pN \rfloor} (\op{i_k}{j_k}+\op{j_k}{i_k})$ in the position basis $\{\ket{i},1\leq i\leq N\}$. The first term describes on-site disorder, with $\varepsilon_i$ i.i.d.~Gaussian random variables with zero mean and standard deviation $W$. The second term runs over nearest neighbors. The third term gives the long-range links that connect pairs $(i_k,j_k)$, randomly chosen with $|i_k-j_k|>1$. The case $p=0$ is the 1D Anderson model. At finite $p$, our system is a random graph with mean connectivity $K=1+2p$, giving access to the regime $1< K\leq 2$.

\emph{Glassy physics approach.---}
We first use a recursive technique used to investigate localization on the Bethe lattice \cite{derrida1988polymers, abou1973selfconsistent, miller1994weak, monthus2008anderson, feigel2010superconductor}. It is exact for a Cayley tree (which has no loop), but only an approximation in the case of a generic graph. For a regular tree with $K+1$ neighbors, the diagonal elements $G_{ii}$ of the Green operator follow $G_{ii}=(\epsilon_i-E-\sum_{j=1}^K G_{jj})^{-1}$, where the sum is over the $K$ children $j$ of node $i$ \cite{abou1973selfconsistent}. In our model, each parent node has either one or two children. This leads to three recursion equations determining the probability distribution of $G$ (see Supp. Mat.).  

\begin{figure}
\includegraphics[width=0.95\linewidth]{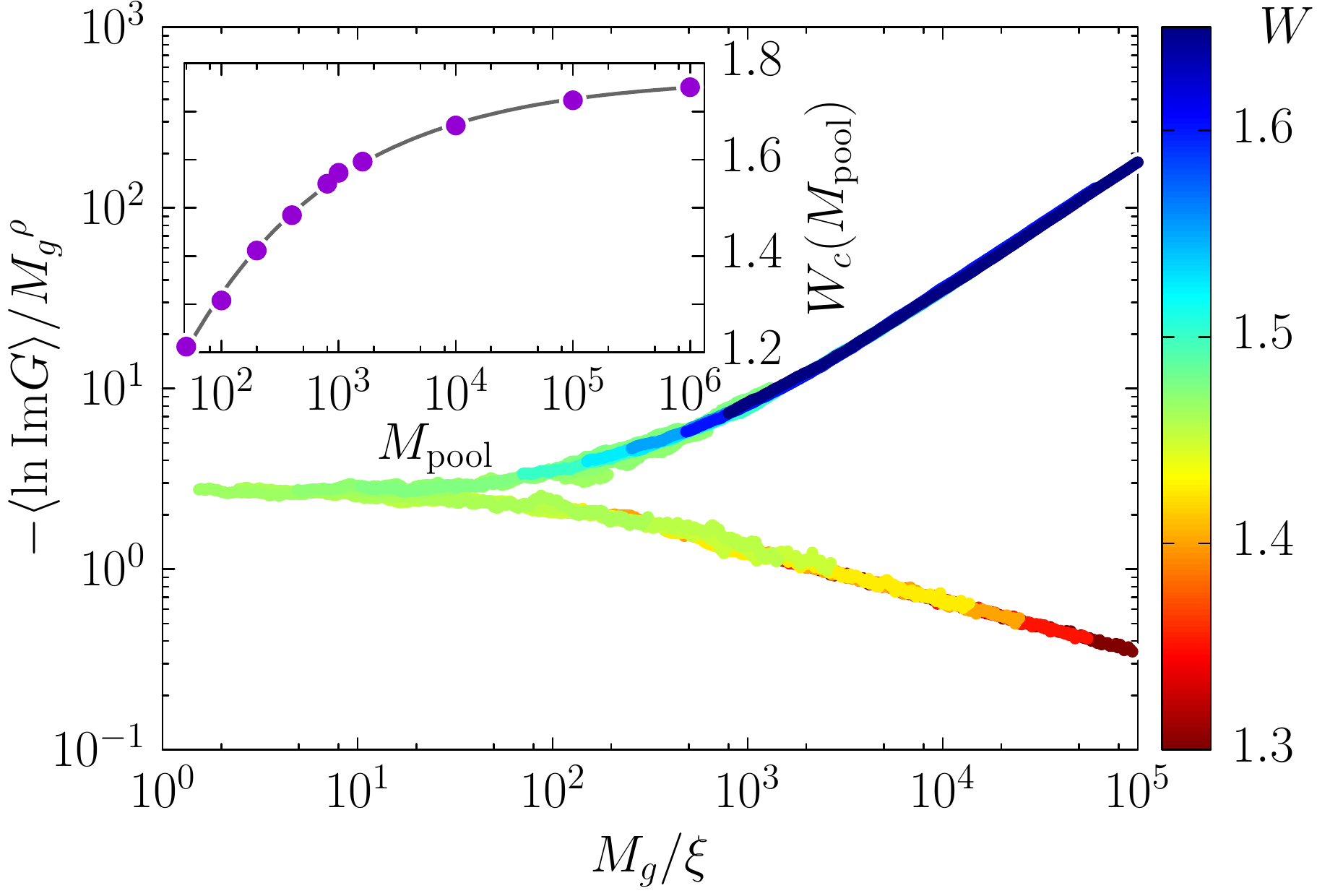}
\caption{Single parameter scaling of $\langle \ln \img\rangle$ versus number of generations $M_g$, for $\mpool=400$, $M_g$ from $2 \mpool$ to $10^5$, $p=0.06$ and $W\in[1.3,1.65]$ (see color code).  $\langle \ln \img\rangle \sim - {M_g}^\rho$ with $\rho \approx 0.28$ at $W=1.46\approx W_c(\mpool)$. Finite-size scaling of $-\langle \ln \img\rangle / {M_g}^\rho$ following Eq.~\eqref{eqscaling}. The scaling parameter $\xi$ diverges at the threshold as $\xi \sim \vert W-W_c(\mpool)\vert^{-\nu}$, with $\nu \approx 1.4$ and $W_c(\mpool)\approx 1.46$. Inset: Value of the critical disorder $W_c(\mpool)$ as a function of $\mpool$. The line is a fit by $W_c(\mpool) = W_c^\infty+A_0 \,{\mpool}^{-\beta}$ with $A_0$ a constant, $W_c^\infty= 1.77$ and $\beta=0.33$. The error bars are of the same order as the fluctuations of the data points.
\label{fig:gabriel} }
\end{figure}

In order to probe the localization properties on the disordered graph, we use the belief propagation method (or pool method), which consists of sampling the distribution of $G$ with a Monte-Carlo approach \cite{miller1994weak, monthus2008anderson}. 
For a fixed value of $E$, we start from an initial pool of $\mpool$ complex values for the local variables $G_{ii}$, $1\leq i \leq \mpool$ and calculate the next generation by applying the recursion relations. 
The important quantity is the typical value of the imaginary part $\img$ which
goes to zero in the localized phase as $\langle \ln \img \rangle\sim -M_g/\xi_l$ when the number of generations $M_g$ tends to infinity, (here $\langle X\rangle$ denotes ensemble averaging)
whereas in the delocalized phase $\langle \ln \img \rangle$ converges to a finite value. 
We observed that the localization length $\xi_l$ diverges at the transition  as $\xi_l \sim \left[W-W_c(\mpool)\right]^{-\nu_l}$ with the critical exponent $\nu_l\approx 1$ and a critical disorder $W_c(\mpool)$ which depends on $\mpool$ (see also \cite{miller1994weak, monthus2008anderson}).
We determined $W_c(\mpool)$ for values of $\mpool$ up to $10^6$. 
The results, presented in the inset of Fig.~\ref{fig:gabriel}, show that $W_c(\mpool)$ converges to $W_c^\infty \approx 1.77$ for $p=0.06$ as $\mpool \to \infty$.

Following \cite{derrida2007survival,monthus2008anderson}, we assume that $\langle \ln \img \rangle$ follows a single parameter scaling law:
\begin{equation}\label{eqscaling}
 \langle \ln \img \rangle = -(M_g)^\rho \; \mathcal F_G(M_g/\xi) \; ,
\end{equation}
with the scaling parameter $\xi \sim \vert W - W_c(\mpool)\vert ^{-\nu}$ \cite{xiglassy}.
In order to sample correctly the distribution of $\img$, values of $\mpool$ as large as possible are usually considered, which entails typically $M_g \leq \mpool$ (see above). However our numerical results show that the scaling behavior \eqref{eqscaling} is valid but visible only for $M_g\gg\mpool$. Moreover, for a given initial pool, the fluctuations of $\langle \ln \img \rangle$ when $M_g$ is varied can be extremely large (especially at criticality). In order to analyze the scaling behavior \eqref{eqscaling} we therefore considered values of $\mpool$ from $50$ to $800$, $M_g$ from $2 \mpool$ to $10^5$, and averaged additionally over $100$ different realizations of the pool. 
The one-parameter scaling hypothesis \eqref{eqscaling} is confirmed by the data collapse shown in Fig.~\ref{fig:gabriel}, which allows us to extract the scaling exponents $\nu\approx 1.4 \pm 0.2$ and $\rho \approx 0.28 \pm 0.07$, that do not depend on $\mpool$. 
Therefore, in the delocalized phase, the typical value of $\img $ vanishes at the transition with an essential singularity 
$\lim_{{M_g \rightarrow \infty}}\langle \ln \img \rangle \sim - (W_c(\mpool) -W)^{-\kappa}$ with $\kappa=\rho \nu \approx 0.39 \pm 0.16$ the critical exponent in the delocalized phase, compatible with the value $1/2$ predicted analytically \cite{mirlin1991universality, *fyodorov1991localization, *fyodorov1992novel, mirlin1994distribution}. Moreover, from $\nu_l=\nu (1-\rho)$ we recover the value $\nu_l \approx 1.0 \pm 0.2$ (see also \cite{monthus2008anderson}).

\begin{figure}
\includegraphics[width=0.95\linewidth]{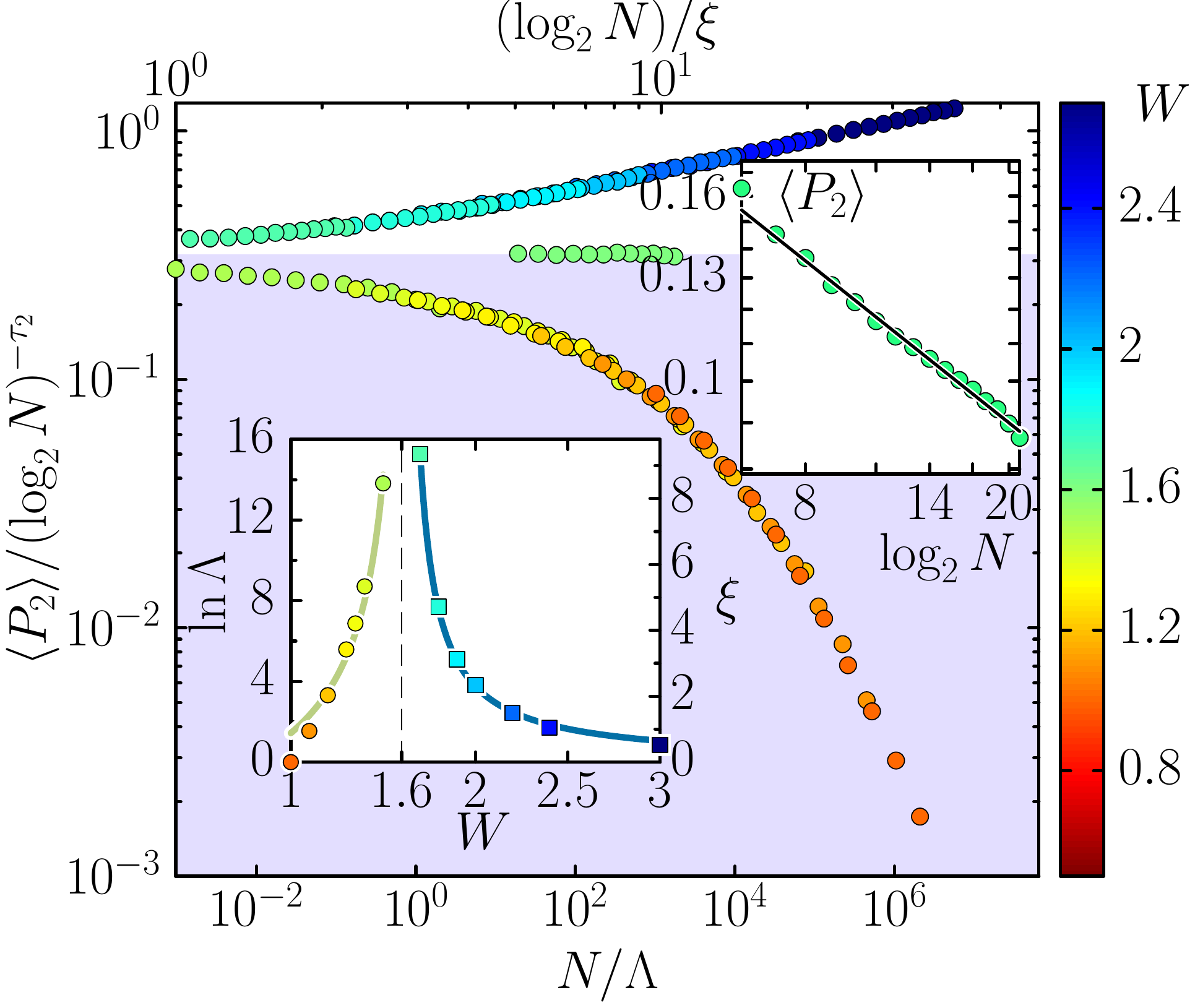}
\caption {Scaling of the moment $\langle P_2\rangle$ with $N$ for $p=0.06$ and $W\in[0.8, 3]$. 
 For $W >W_c$ (localized phase), linear scaling of $\langle P_2\rangle / (\log_2 N)^{-\tau_2}$ following Eq.~\eqref{eq:scalin} with $\tau_2=0.42$. For $W<W_c$ (delocalized phase), volumic scaling following Eq.~\eqref{eq:scavol}. Data for $W=W_c$ have been shifted horizontally for visibility. Upper inset (note the log-scale): The critical behavior of $\langle P_2\rangle$ at $W=1.6\approx W_c$ is very well fitted (black line) by  $\langle P_2\rangle =A_0 (\log_2 N)^{-\tau_2}$ with $A_0$ a constant and $\tau_2 \approx 0.42$. Lower inset: correlation volume $\Lambda$ (circles) and localization length $\xi$ (squares) on both sides of the transition. Solid lines are the fits: $\ln \Lambda = A_1 + A_2 (W_c- W)^{- \kappa}$ and $\xi=A_3 (W-W_c)^{-\nu_l}$, with $W_c=1.6$ (assigned value), yielding $\kappa\approx 0.46$ and $\nu_l \approx 1$. Error bars on $\langle P_2 \rangle$ are below symbol size, and not taken into account in the scaling analysis.
\label{fig:scalingP2} }
\end{figure}

\emph{Scaling analysis of eigenfunction moments.---}
We now describe the results of our second approach. We performed exact diagonalizations of a large number of realizations of graphs with up to $N\sim 2\times 10^6$ sites and obtained for each realization $16$ eigenfunctions closest to the center of the band using the Jacobi-Davidson iterative method~\cite{JacobiDavidson}. We performed a multifractal analysis of the eigenfunctions $\vert \psi \rangle$ 
by considering the scaling of average moments $\langle P_q \rangle =\langle\sum_{i=1}^N |\psi_i|^{2 q}\rangle$ for real $q$ as a function of $N$.
For a $d$-dimensional system of linear size $L$ and volume $N=L^d$, multifractal eigenfunctions have $\langle P_q \rangle\sim L^{-\tau_q}$ at large $L$, or  equivalently $\langle P_q \rangle\sim N^{-\chi_q}$ with $\chi_q=\tau_q/d$, defining non-trivial multifractal dimensions $D_q=\tau_q/(q-1)$. In the localized case $D_q=0$ whereas for  wavefunctions delocalized over the whole space $D_q=d$. Our graphs however correspond to a case of infinite dimensionality \cite{Dorogovtsev2010}, where the system volume $N=V_g(d_N)$ is exponential in the linear size $d_N\sim \log_2 N$, the diameter of the system (see Supp. Mat.).  Figure \ref{fig:scalingP2} shows that a critical behavior $\langle P_2 \rangle  \sim (\log_2 N)^{-\tau_2}\sim {d_N}^{-\tau_2}$ actually holds with $\tau_2\approx 0.42$ for $p=0.06$ and $W=1.6\approx W_c$ (upper right inset). This ${d_N}^{-\tau_q}$ dependence entails that a behavior $\langle P_q \rangle\sim N^{-\chi_q}$ would lead to $\chi_q=0$, in line with the analytical predictions \cite{castellani1986upper,mirlin1994distribution,evers2008anderson} at infinite dimensionality. Moreover, in the light of the analogy with directed polymers, in the localized phase eigenfunctions are located on few branches of the graph on which they are exponentially localized. At criticality the localization length diverges to the system size $ d_N$ and one should observe critical wave functions located on few branches on which they display additional multifractal fluctuations.

The following one-parameter scaling hypothesis should naturally follow:
\begin{equation} \label{eq:scalin}
 \langle P_q \rangle = d_N^{- \tau_q} \mathcal F_{\mathrm{lin}} (d_N/\xi) \; .
\end{equation}
It is consistent with the scaling theory for both the AT in finite dimension \cite{rodriguez2011multifractal} and the glassy physics approach detailed above. A careful finite-size scaling analysis of our data shows that \eqref{eq:scalin} yields a very good data collapse on the localized side of the transition, see Fig.~\ref{fig:scalingP2}, upper branch in the main panel. The scaling parameter $\xi\sim \xi_l$, with $\xi_l$ the localization length, diverges as $\xi\propto (W-W_c)^{-\nu_l}$ near the AT, with $\nu_l \approx 1. \pm 0.1$ (in agreement with the value found by our first glassy physics approach). 

However, in the delocalized phase, small but systematic deviations are observed (see Supp. Mat.). This leads us to propose a different scaling in this phase. Indeed, the linear scaling \eqref{eq:scalin} is not the only possibility:
the system volume $N$ could instead be rescaled by a characteristic volume $\Lambda$:
\begin{equation} \label{eq:scavol}
\langle P_q \rangle = d_N^{- \tau_q} \mathcal F_{\mathrm{vol}} (N/\Lambda).
\end{equation}
Both scaling hypotheses \eqref{eq:scalin} and \eqref{eq:scavol} are strictly equivalent in finite dimension, but lead to very different behaviors for a graph of infinite dimensionality.
In the first linear scaling picture \eqref{eq:scalin}, 
the delocalized states consist of the repetition of linear critical structures of size $ \xi $ and the moments behave as $\langle P_q \rangle \approx  \xi^{- \tau_q}  N^{-(q-1)/\xi}$. It is reminiscent of the non-ergodic behavior discussed in \cite{deluca2014anderson,kravtsov2015random, altshuler2016non, facoetti2016non}. In the second volumic scaling picture \eqref{eq:scavol}, a delocalized state consists of $N/\Lambda$ volumic critical structures of size $\Lambda$, and moments behave as $\langle P_q \rangle \approx \left(\frac{\Lambda}{N}\right)^{q-1} \left(1-\tau_q \frac{\xi}{d_N}\right)$ (see Supp. Mat.). This is consistent with previous analytical results \cite{mirlin1991universality, *fyodorov1991localization, *fyodorov1992novel, mirlin1994distribution}.
The finite size scaling shown in Fig.~\ref{fig:scalingP2} clearly indicates that the volumic scaling \eqref{eq:scavol} puts all the curves onto a single scaling function in the delocalized phase $W<W_c$,  with correlation volume diverging exponentially at the transition as $\ln \Lambda \approx (W_c - W)^{-\kappa}$,  $\kappa\approx 0.46 \pm 0.1$ (in good agreement with our glassy physics approach,  the analytical prediction $\kappa=1/2$ \cite{mirlin1991universality, *fyodorov1991localization, *fyodorov1992novel, mirlin1994distribution} and the recent numerical results \cite{tikhonov2016anderson}).

\begin{figure}
\includegraphics[width=0.95\linewidth]{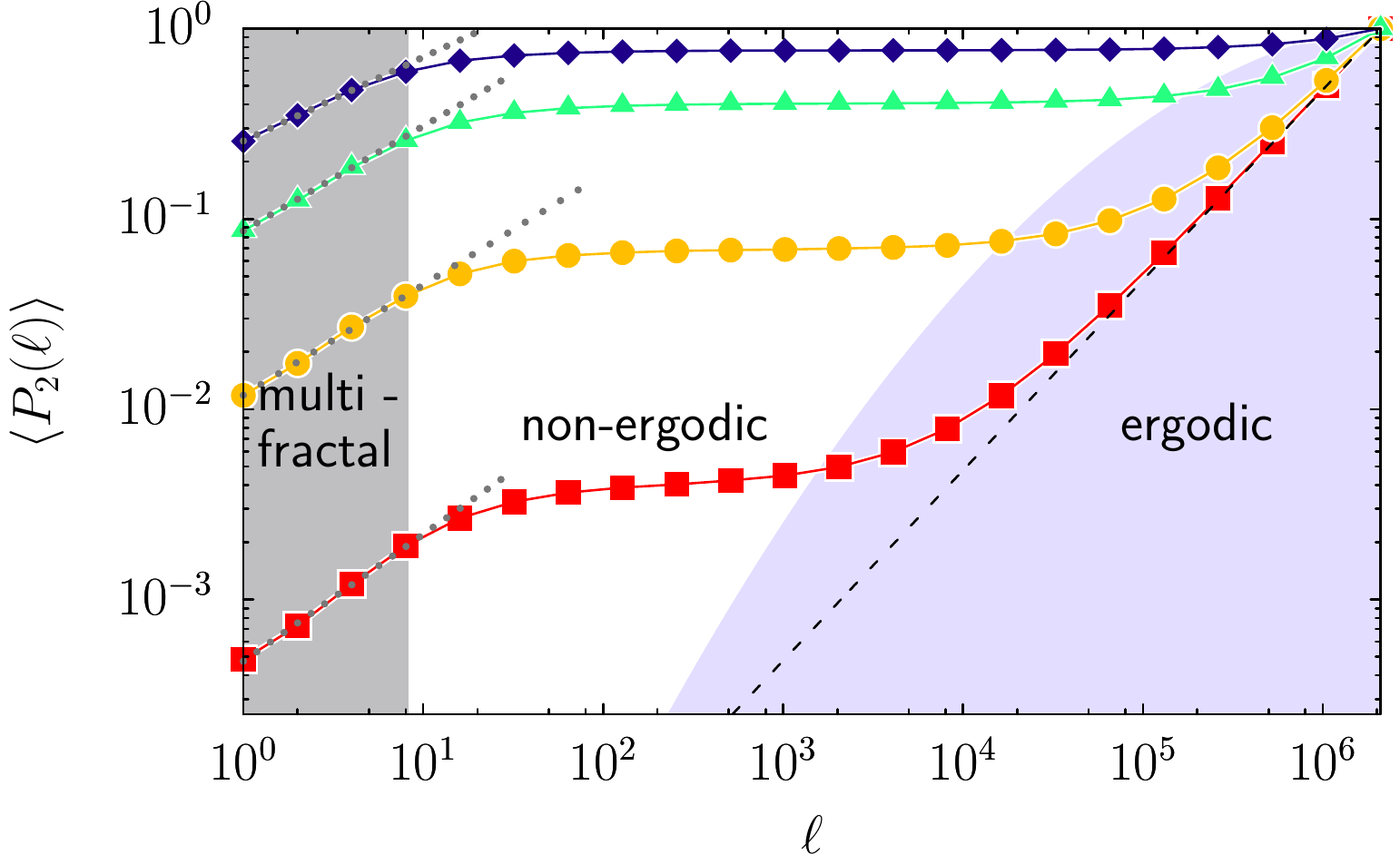}  
\caption{ $\langle P_2\rangle$ versus the box volume $\ell$ for $N=2^{21}$ with $p=0.06$ and $W=1.0\, (\square),\ 1.3\, (\circ),\ 1.6\, (\triangle),\  2.4\, (\diamond) $ . The gray shaded area on the left delimits the small $\ell<1/2p$ multifractal behavior (dotted lines have slope ranging from 0.44 to 0.66). The light-blue shaded area on the right delimits the large scale $\ell \gg \nev$ ergodic behavior (dashed line has slope 1).
\label{fig:P2deel}}
\end{figure}

\begin{figure}
\includegraphics[width=0.95\linewidth]{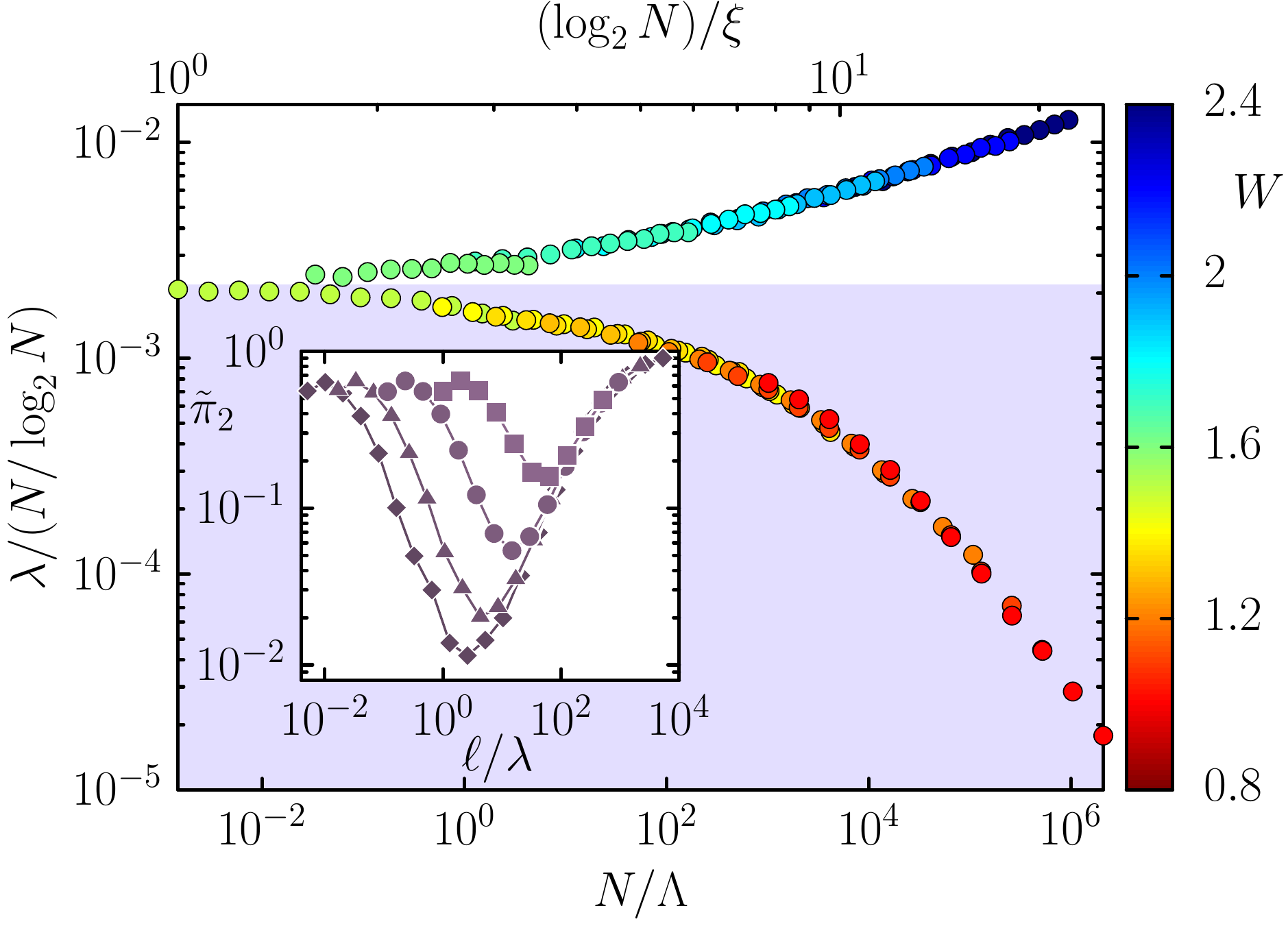} 
\caption{Scaling behavior of the NEV $\nev$ with $N=2^{10}$ to $2^{21}$, $p=0.06$ and $W\in[0.8, 2.4]$. At the threshold $W=W_c\approx 1.6$, $\nev \sim N/\log_2 N$. In the localized phase, the data collapse is excellent when plotted as a function of $(\log_2 N)/\xi$ (linear scaling, see text). In the delocalized phase, a volumic scaling as function of $N/\Lambda$ makes the curves collapse. Inset: Determination of the NEV $\nev$ through rescaling of $\tilde{\pi}_2(\ell)$ data (see text) versus $\ell/\nev$, in the large scale $\ell\gg\nev$ ergodic regime, for $W=1.6$ and $N=2^{10}\, (\square),\, 2^{14}\, (\circ),\, 2^{18}\, (\triangle)$ and $2^{21}\, (\diamond)$.  
\label{fig:scalingXI_p006}}
\end{figure}

\emph{Non-ergodicity volume.---}
In order to probe the local properties of localization in our system, we use the box-counting method, which consists of investigating the scaling properties of moments of coarse-grained wavefunctions. Dividing the system of $N$ sites into boxes of $\ell$ consecutive sites along the lattice (i.e.\ not following the long-range links) and defining a measure $\mu_k=\sum_{i\in\textrm{box }k}|\psi_i|^2$ of each box, moments are defined as $P_q(\ell)=\sum_k \mu_k^q$. For a multifractal state, they are expected to scale as $\langle P_q(\ell)\rangle\sim \ell^{\pi_q}$ at large $N$ with nontrivial $\pi_q$ \cite{RemyPRL2014,*RemyPRE2015}. In the localized case $\pi_q=0$ whereas for a wavefunction delocalized over the whole system $\pi_q=q-1$. Figure \ref{fig:P2deel} displays the moments $\langle P_2(\ell)\rangle$ as a function of $\ell$ for different $W$, and shows that three distinct regimes can be identified.   At scales below the mean distance  $1/(2p)$ between two long-range links, moments have a power-law behavior with $\pi_2\approx 0.5 \pm 0.1$, in the vicinity of $W_c$ where $\xi \gg 1/(2p)$. For $\ell \leq 1/(2p)$, one is probing only one branch, therefore the value of $\pi_2$ can be seen as a measure of the critical multifractality on few branches.  $\pi_2$ is indeed close to $\tau_2\approx 0.42$ found above. At intermediate scales, the moments follow a plateau characteristic of a strongly non-ergodic behavior. This regime corresponds to the critical behavior we observe when changing $N$, i.e. $\langle P_2 \rangle \sim (\log_2 N)^{-\tau_2}$, thus $\langle P_2 \rangle \sim N^{-\chi_2}$ with $\chi_2=0$ (see Fig. \ref{fig:scalingP2} and Fig. S4 of Sup. Mat.). Beyond a certain characteristic scale $\nev$ which depends on $W$ and $N$, the moments are linear in $\ell$, which corresponds to an ergodic behavior.  In the localized case $\nev\sim N$, so that there is no ergodic behavior, while in the delocalized case, $\nev$ saturates to a finite value (see below), and states are ergodic at scales above $\nev$, non-ergodic below: we therefore call $\nev$ the NEV. One can extract $\nev$ from a rescaling of the local slopes $\tilde{\pi}_q(\ell) \equiv \frac{d \ln \langle P_q(\ell) \rangle}{d\ln \ell}$ \cite{RemyPRL2014, *RemyPRE2015} in the ergodic regime $\ell \gg \nev$ (see inset of Fig.~\ref{fig:scalingXI_p006}). 

 The data shown in Fig.~\ref{fig:scalingXI_p006} can be described by a linear scaling $\frac{\nev}{N/\log_2 N} = \mathcal G_\mathrm{lin} \left( \frac{\log_2 N}{\xi} \right)$ in the localized regime, with $\xi\sim \xi_l$ the localization length, and by a volumic scaling $\frac{\nev}{N/\log_2 N} = \mathcal G_\mathrm{vol} \left( \frac{N}{\Lambda} \right)$ in the delocalized regime. At the threshold, the critical behavior $\lambda \sim N /\log_2 N$ shows that states are located on few branches of $\log_2 N$ sites. This confirms the description of the critical wave functions as having multifractal fluctuations on a logarithmically small fraction of the system
volume. Moreover, the volumic scaling shows that in the limit $N\gg \Lambda$ the NEV $\lambda$ saturates to the correlation volume $\Lambda$ (see Supp. Mat.). This implies that the delocalized phase is ergodic in the limit of large $N\gg \Lambda$.

We have checked that the scaling properties are the same for $q\geq 1 $ whereas for $ q < 1$ the volumic behavior of the delocalized phase extends to the critical and localized regimes. This is to be expected \cite{mirlin1994distribution, monthus2011anderson}: for $q<1$ all small values of the wavefunction, even outside the few localization branches, contribute to the moment.

\emph{Universality.---}
We checked the universality of our results by considering different values of the graph parameter from $p=0.01$ to $ p = 0.49$ (see Supp. Mat.), which changes the average branching parameter $K=1+2p$ considerably from $K=1.02$ to $K=1.98$. Our data show that the critical scalings are insensitive to the value of $p$. Moreover, the critical exponents have universal values $ \kappa \approx  0.5$ and $ \nu_l \approx 1$.

\emph{Conclusion.---}
Our study strongly supports the following picture:
A single transition separates a localized phase from an ergodic delocalized phase. In the delocalized phase, the NEV $\Lambda$ marks the threshold between a non-ergodic behavior reminiscent of glassy physics at small scales and an ergodic behavior at large scales. At the transition, $\Lambda$ diverges, so that the behavior below $\Lambda$ extends to the whole system. This highlights a new type of strong non-ergodicity with states located on few branches on which they display additional multifractal fluctuations. The ergodic character of the delocalized phase is controlled by the unusual volumic scaling in this phase, different from the linear scaling which applies in the localized phase and to the whole transition in the Cayley tree as found in our glassy physics approach. Therefore the absence of boundary may change the nature of the AT, as envisioned in \cite{tikhonov2016fractality}. 

Our results apply to random graphs up to $K=2$, for which they are in agreement with \cite{mirlin1991universality, *fyodorov1991localization, *fyodorov1992novel, mirlin1994distribution, tikhonov2016anderson, biroliprivate}. 
Recent results \cite{altshuler2016non,altshuler2016multifractal} suggest that a non-ergodic delocalized phase could still arise at large $K$ (see however \cite{facoetti2016levy}). It will thus be very interesting to investigate this regime $K \gg 2$ with our approach. Last, non-ergodicity is usually linked with a specific dynamics, such as anomalous diffusion which will be instructive to study. Another fascinating perspective would be to probe if our non-trivial scaling theory applies to many-body localization.

\acknowledgments{We thank Claudio Castellani and Nicolas Laflorencie for interesting discussions. We thank CalMiP for access to its supercomputer and the Consortium des \'Equipements de Calcul Intensif (C\'ECI), funded by the Fonds de la Recherche Scientifique de Belgique (F.R.S.-FNRS) under Grant No. 2.5020.11.
This work was supported by Programme Investissements
d'Avenir under the program ANR-11-IDEX-0002-02, reference  ANR-10-LABX-0037-NEXT,  by  the  ANR  grant
K-BEC  No  ANR-13-BS04-0001-01,  by  the  ARC  grant
QUANDROPS 12/17-02 and by the CONICET- CNRS
bilateral project PICS06303R.}
 
%



\onecolumngrid
\newpage
\appendix
\begin{center}
{\large\textbf{Supplemental material to\\  ``Scaling theory of the Anderson transition in random graphs: \\ergodicity and universality''
}}
\end{center}
\setcounter{equation}{0}
\setcounter{figure}{0}
\setcounter{table}{0}
\setcounter{page}{1}
\makeatletter
\renewcommand{\theequation}{S\arabic{equation}}
\renewcommand{\thefigure}{S\arabic{figure}}
\renewcommand{\bibnumfmt}[1]{[S#1]}
\renewcommand{\citenumfont}[1]{S#1}

\subsection*{Topological properties of the random graph model}

In this Section we want to assess the topological properties of the random graphs considered in our model more in details. In particular random graphs often refer to Erd\"os-Renyi random graphs, i.e. graphs having $N$ vertices and $(p+1)N$ links which are chosen with a uniform probability over the set of all such graphs. In this Section it will be explained that, while the set of random graphs considered here is a subset of the set of Erd\"os-Renyi random graphs, they share several generic features of those graphs. This strenghtens the idea that our model for graphs posesses generic properties. We also want to stress that our random graph model coincides with the model of Random Regular Graphs (RRG) at the limit $p=0.5$.

\begin{figure}[h]
\includegraphics[width=0.5\linewidth]{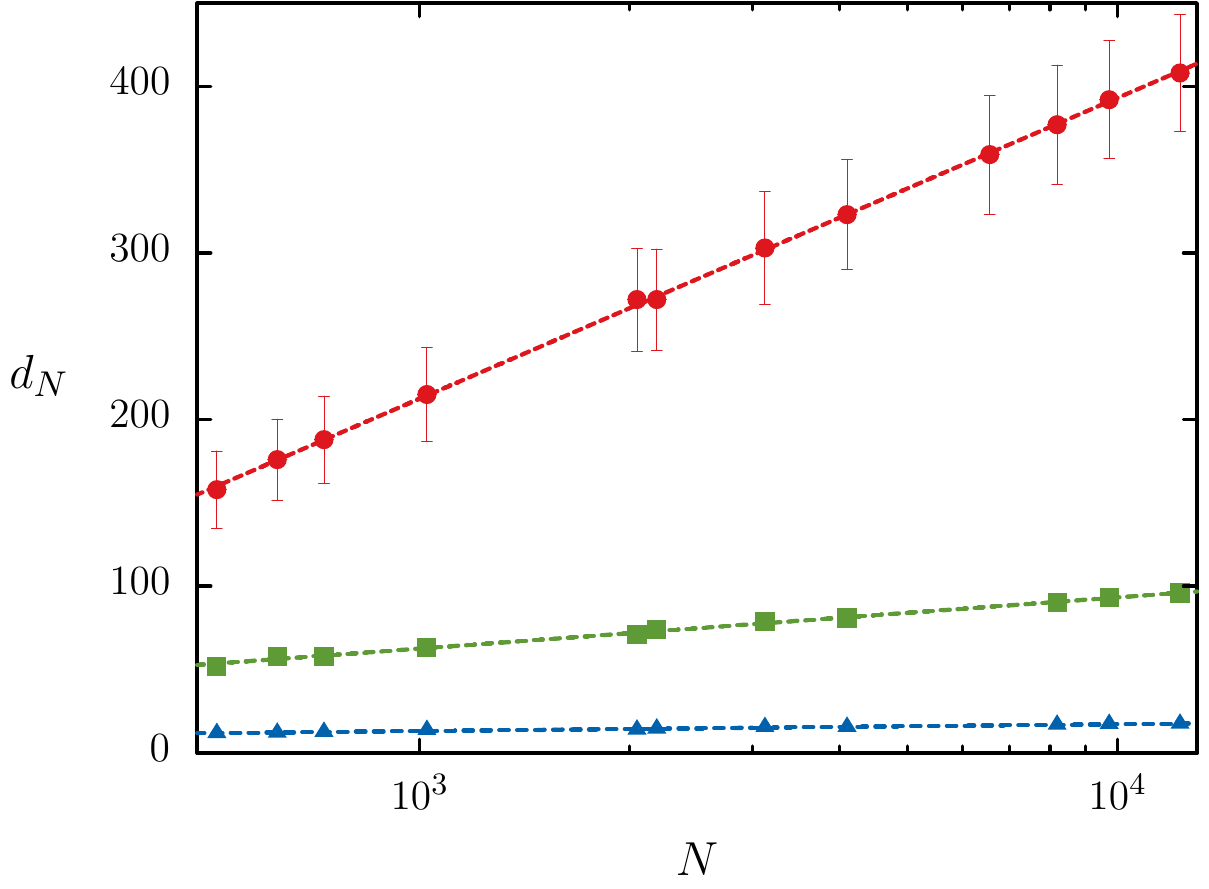}
\caption{(Color online) Diameter $d_N$ of our random graph model as a function of the total number $N$ of vertices. For each value of $N$, $10000$ random graphs were sampled. The values reported here are the center of a Gaussian fit of the distribution of their diameter. Red circles: numerical data for $p=0.01$. Red dashed line: Logarithmic fit, $d_N\sim 78 \ln (0.015 N)$. Green squares: numerical data for $p=0.06$. Green dashed line: Logarithmic fit, $d_N\sim 13 \ln (0.10 N)$. Blue triangles: numerical data for $p=0.49$. Blue dashed line: Logarithmic fit, $d_N\sim 1.8\ln(1.5 N)$.}
\label{diameter_vs_N}
\end{figure}

The first reason why the considered random graphs are generic, is that they have infinite dimension. The dimension refers here to the Hausdorff dimension of a graph. First one defines a distance between two vertices $i$ and $j$ as the number of vertices of a shortest path connecting $i$ and $j$. Then one can compute the mean pair distance $\overline{l}$ as the average distance between any pair of vertices of the graph. Another quantity of interest for our study is the diameter, or the linear size, of our graph. It is denoted by $d_N$ and is defined as the largest distance between any pair of vertices of the graph.
The Hausdorff dimension of the graph is evaluated by taking a sequence of graphs with increasing number of vertices $N$. The variation of $\overline{l}$ as a function of $N$ leads to the definition of the Hausdorff dimension $d_{\rm H}$, see e.g.~\cite{S-Dorogovtsev2010}:
\begin{equation*}
  \overline{l}\sim N^{1/d_{\rm H}},\ N\to\infty\ .
\end{equation*}
Erd\"os-Renyi random graphs have infinite dimension as the mean pair distance grows like $\overline{l}\sim\ln N$ for large $N$. We checked that this is also the case for our model of random graphs whenever $p$ is positive. More precisely for $p=0.49$ we found numerically an asymptotic form $\overline{l}\approx 1.44\ln N$, which agrees with the prediction $\overline{l}\approx \ln N/\ln 2$ for $3-$regular graphs in RRG.
Such a scaling for the mean pair distance also means that the diameter grows logarithmically as a function of $N$ for large graphs. This is illustrated for three different values of $p$ in Fig.~\ref{diameter_vs_N}. In particular for $p=0.49$ the best fit is close to the prediction 
$d_N\approx\ln N/\ln 2$ for $3-$regular graphs in RRG (see Theorem 3 in \cite{S-bollobas1982diameter}).

The second common feature between our graphs and Erd\"os-Renyi random graphs is the clustering coefficient. It is defined as the probability that two given neighbors of a fixed vertex are themselved connected by a link. An important property of Erd\"os-Renyi random graphs is that, when the vertex number $N$ increases, the clustering coefficient grows like $N^{-1}$ \cite{S-Dorogovtsev2010}. For our model of random graphs, it was found numerically that the clustering coefficient obeys the following law, for any $p$:
\begin{equation}
  \label{fitC_vs_p}
  C(N,p)\simeq \frac{3}{N}p(1-p)\ ,
\end{equation}
which agrees with the scaling for Erd\"os-Renyi random graphs.

Another common point with Erd\"os-Renyi random graphs is that the graphs considered in our study locally look like a tree. More precisely the number of small loops is independent of the number of vertices~\cite{S-Dorogovtsev2010}.
This is another common point with Erd\"os-Renyi random graphs.

Eventually we want to stress that our model of random graphs share also the main features of a smallworld network \cite{S-Str00}. This connection is justified by looking at the mean pair distance as a function of $p$. Indeed, we checked that $\overline{l}$ quickly decreases as a function of $p$.

\subsection{Recursion equations}
 For a regular tree graph with $K+1$ neighbors, the recursion equation is obtained by considering the Green operator $G(E)=(M-E\, \mathbb{I})^{-1}$ of the adjacency matrix $M$ of the network with some vertex $m$ removed. If $i$ is a child node of $m$, the diagonal entry $G_{ii}(E)$ of the Green function can be expanded (see e.g.~\cite{S-bogomolny2013calculation}) as $G_{ii}=(M_{ii}-E-\sum_{j=1}^K G_{jj})^{-1}$, where the sum runs over the neighbors $j$ of node $i$ other than $m$. As the tree is self-similar, the $G_{jj}$ and $G_{ii}$ all have the same probability distribution $P(G)$. Moreover, the $G_{jj}$ are independent and also are independent of the random variables $M_{ii}=\epsilon_i$, so that at a fixed value of the energy $E$ the above relation determines a functional equation for the probability $P(G)$. 

In order to obtain a similar recursion relation for our random graph model, we consider the local tree-like structure of the graph. The tree is such that each parent node has either one or two children, with probability respectively $1-2p$ and $2p$. Following the cavity method \cite{S-mezard1990spin}, we consider a graph where some parent node (say $m$) has been removed. Each child node $i$ is the root of a tree which can be of three different types: either $i$ has one remaining neighbor (case A), or two neighbors connected by two nearest-neighbor links (case B), or two neighbors connected by one nearest-neighbor and one long-range link (case C). We thus have to distinguish between three types $A_i$, $B_i$ and $C_i$, of local random variables $\tilde{G}_{ii}(E)$, corresponding to the three possible local patterns. We introduce a fourth type G of random variable $G_i$, equal to $A_i$ with probability $1-2p$ and to $B_i$ with probability $2p$. Local pattern of type A are those where node $i$ has a single child $j$, which can itself be of type A or B (note that 
type C is excluded since the link between $i$ and $j$ is of nearest-neighbor type). That is, the child is of type G. In case B, node $i$ has a neighbor of type C and a neighbor of type G. In case C, the two children are of type G. The analog of the recursion equation on the regular tree now takes the form of three recursion relations
\begin{eqnarray}
\label{recursionABC}
A_i&=&\frac{1}{\epsilon_i-E- G_j},\\
B_i&=&\frac{1}{\epsilon_i-E-G_{j_1}-C_{j_2}},\\
C_i&=&\frac{1}{\epsilon_i-E-G_{j_1}-G_{j_2}},
\end{eqnarray}
together with the condition
\begin{equation}
\label{recursionDAB}
G_i=\left|\begin{array}{cl}
A_i&\quad\textrm{ with probability }1-2p\\
B_i&\quad\textrm{ with probability }2p
\end{array}\right.\, .
\end{equation}
The probability distributions for each type of random variable follow a set of functional equations that can be directly inferred from Eqs.~\eqref{recursionABC}--\eqref{recursionDAB}.

\subsection{Scaling analysis}

\subsubsection{Deviations to the linear scaling hypothesis in the delocalized regime}

\begin{figure}
  \includegraphics[width=0.4\textwidth]{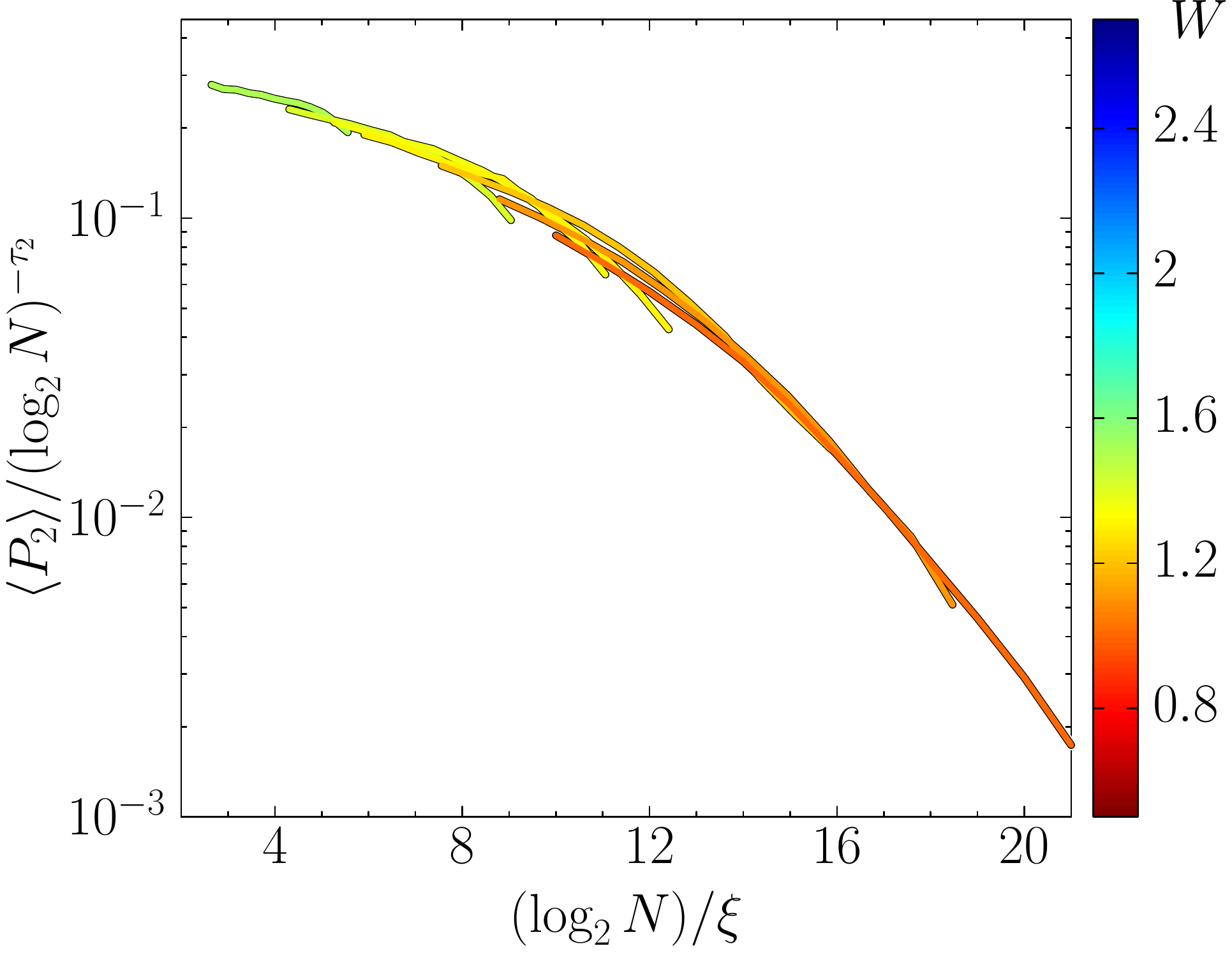}
  \caption{Deviations to the linear scaling hypothesis \eqref{eq:scalin} for the moment $\langle P_2 \rangle$ in the delocalized regime $W<W_c\approx 1.6$ for $p=0.06$. The best rescaling of the data as a function of $(\log_2 N)/\xi$ shows small but systematic deviations. The data are represented by lines of different colors according to $W$.}
  \label{fig:DevSupMat}
\end{figure}

Here we show the results of finite-size scaling of the moment $\langle P_2 \rangle$ for $p=0.06$ following the linear scaling hypothesis \eqref{eq:scalin} in the delocalized regime $W<W_c\approx 1.6$. In Fig.~\ref{fig:DevSupMat}, the best rescaling of the data when the linear size of the graph $d_N\sim \log_2 N$ is rescaled by the scaling parameter $\xi$ is shown. Small but systematic deviations are observed (data have been represented with lines to better see these deviations). Clearly, the lines corresponding to different values of the disorder strength do not have the right curvature to be put on each other via such a rescaling. On the contrary, we recall that the linear scaling hypothesis \eqref{eq:scalin} is fully consistent with our data in the localized regime $W>W_c$ (see Fig.~\ref{fig:scalingP2}).

We stress that the systematic deviations observed in Fig.~\ref{fig:DevSupMat} cannot be accounted for by irrelevant corrections. Irrelevant corrections (which have been shown to be important in scaling analysis, as e.g. for the Anderson transition in 3D \cite{S-Slevin99}) are important in the vicinity of the critical point and are less and less significant as one goes away from the critical point: indeed they are corrections to the critical behavior and thus their change is much weaker than the effect of the distance to the critical point.
Instead, the systematic deviations we observe do not decrease substantially for $\langle P_2 \rangle$ varying over two orders of magnitude away from the critical point. Moreover in our scaling analysis we tested different assumptions for the critical behavior and the deviations from the volumic scaling were always observed (data not shown).  

\subsubsection{Universality versus a change of the graph parameter $p$}

\begin{figure}
\quad
  \includegraphics[width=0.4\textwidth]{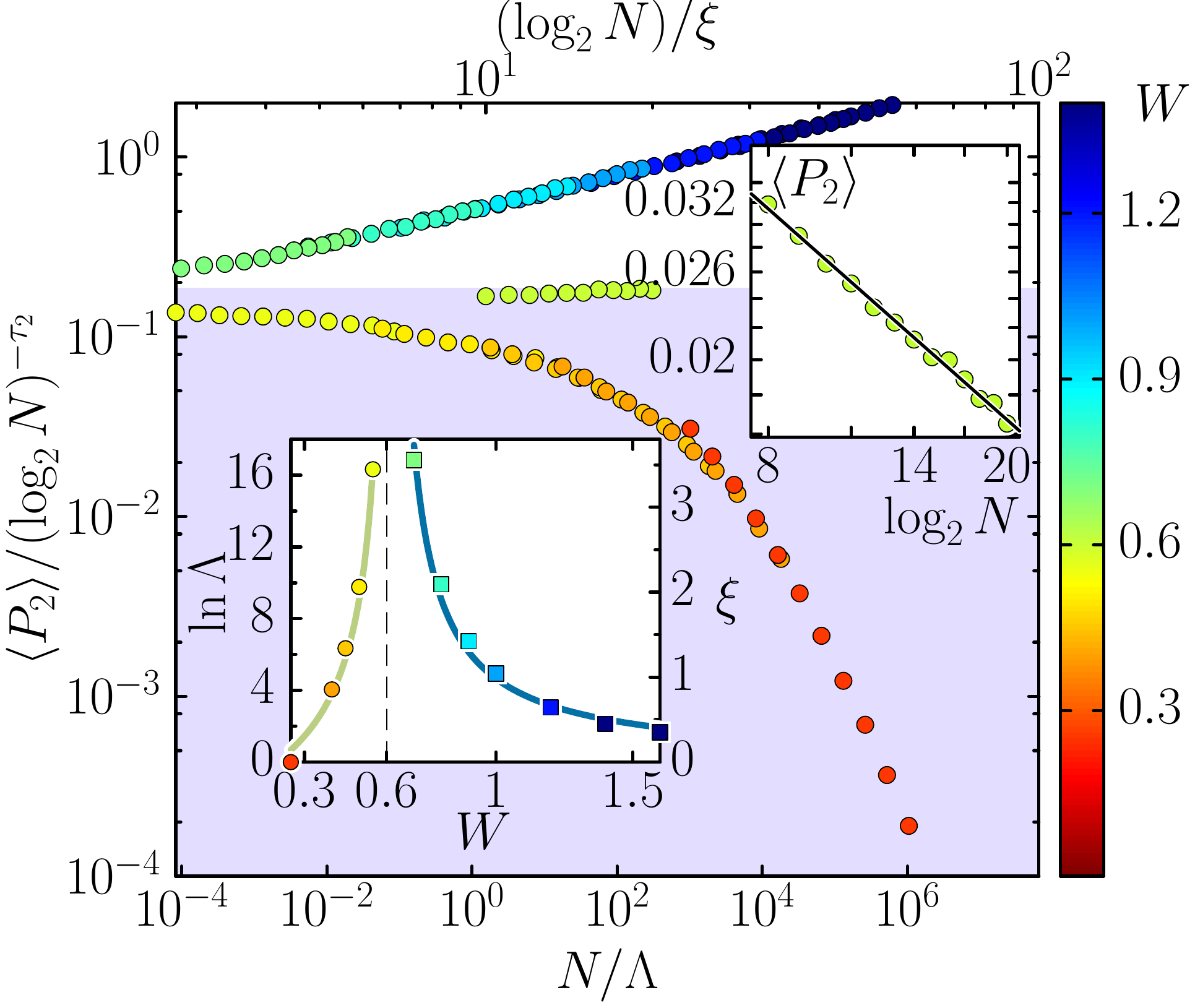}
  \hskip 0.5 cm
   \includegraphics[width=0.4\textwidth]{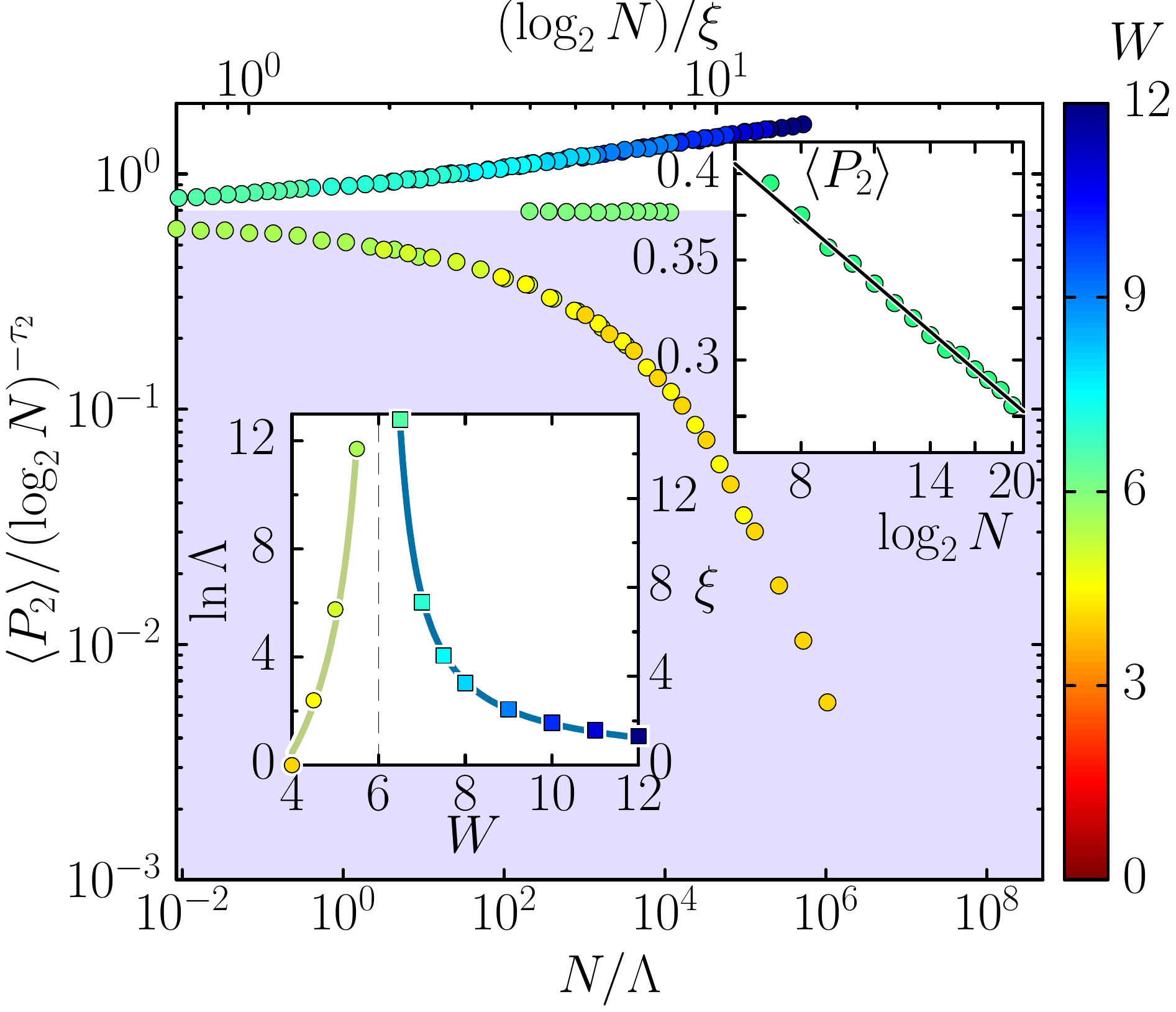}
  \quad
  \caption{Scaling of the moment $\langle P_2\rangle$ for the graph model with $N$ for two values of $p$.  Data for $W=W_c$ have been shifted horizontally for visibility.
  (Left panel) $p=0.01$, $W\in [0.25,1.6]$. At $W= 0.6 \approx W_c$,  the behavior of $\langle P_2\rangle$ is well fitted by $ \langle P_2\rangle = A_0 (\log_2 N)^{-\tau_2}$ with $A_0$ a constant and $\tau_2 = 0.69$ (upper inset). In the main panel, the value $\tau_2=0.8$ was considered because it provides the best global scaling. In the localized regime,
  a linear scaling \eqref{eq:scalin} as a function of $(\log_2 N)/\xi$ gives a very good collapse of all the curves for different disorder strengths $W>W_c$. In the delocalized regime, the rescaling of the volume $N$ by the correlation volume $\Lambda$ \eqref{eq:scavol} puts all the data onto each other. Lower inset: correlation volume $\Lambda$ (circles) and localization length $\xi$ (squares) on both sides of the transition. Solid lines are the fits: $\log_2 \Lambda = A_1 + A_2 (W_c- W)^{- \kappa}$ and $\xi=A_3 (W-W_c)^{-\nu_l}$, with $W_c=0.6$ (assigned value), yielding $\kappa= 0.48$ and $\nu_l =0.96$.
 (Right Panel) $p=0.49$, $W\in [3,12]$.  At $W= 6 \approx W_c$,  $\langle P_2\rangle \sim (\log_2 N)^{-\tau_2}$ with $\tau_2 \approx 0.306$ (upper inset). In the main panel the value $\tau_2=0.3$ gives the best global scaling. The same scaling procedure as for $p=0.01$ and $0.06$ has been followed here. Lower inset: Solid lines are the fits: $\log_2 \Lambda = A_4 + A_5 (W_c- W)^{- \kappa}$ and $\xi=A_6 (W-W_c)^{-\nu_l}$, with $W_c=6$ and $\kappa= 0.5$ (assigned values), yielding $\nu_l =1.0$. }
 \label{fig:UnipSupMat}
\end{figure}

Here we present the results of the same scaling analysis represented in Fig.~\ref{fig:scalingP2} for two distinct values of the graph parameter, $p=0.01$ and $p=0.49$.
The same critical scalings and critical exponents are observed in Fig.~\ref{fig:UnipSupMat}: the linear scaling law \eqref{eq:scalin} where the linear system size $d_N\sim \log_2 N$ is rescaled by the scaling parameter $\xi$ describes the data in the localized regime $W>W_c$ while a volumic scaling law \eqref{eq:scalin} as a function of the ratio of the system volume $N$ and the correlation volume $\Lambda$ holds in the delocalized regime. As in the case $p=0.06$ shown in the paper (see Fig.~\ref{fig:scalingP2}), the scaling parameter $\xi\sim\xi_l$, with $\xi_l$ the localization length, diverges at the transition point as $\xi \sim (W-W_c)^{-\nu_l}$ with $\nu_l\approx 1$, while the correlation volume $\Lambda$ diverges exponentially as $\log_2 \Lambda \sim (W_c- W)^{- \kappa}$ with the critical exponent $\kappa\approx 0.5$. 

\subsubsection{Asymptotic behaviors in the delocalized phase}
The linear scaling hypothesis \eqref{eq:scalin} predicts a non-ergodic delocalized phase in the sense of \cite{S-deluca2014anderson,S-kravtsov2015random} which has its origin in the glassy non-ergodicity of the localized phase. Following a scaling law hypothesis, a delocalized state is built from critical structures of size $ \xi $. As a critical structure consists of few branches, a delocalized state is to follow these branches for $ \xi $ steps, then to make connections to $ K $ branches, then follow the $ K $ branches for $ \xi $ steps that perform each $ K $ connections, etc. One thus sees that the number of critical structures constituting a delocalized state is $V_g(d_N/\xi)$ where $V_g(X)$ is the volume of a graph of size $X$. Therefore, the asymptotic behavior of $P_q$ in the delocalized regime should be: $\langle P_q \rangle \sim \xi^{- \tau_q}/{V_g(d_N/\xi)}^{q-1}$ where the number $V_g(d_N/\xi)$ appears on the denominator due to normalization. Because the volume $V_g(X)$ of graphs of infinite dimensionality scales exponentially with the linear size $X$, 
\begin{equation}
\langle P_q \rangle \approx  \xi^{- \tau_q}  N^{-(q-1)/\xi}
\end{equation}
 in the delocalized phase, $d_N \gg \xi$. This behavior corresponds to the following asymptotic dependence of the scaling function $\mathcal F_{\mathrm{lin}}$ (see \eqref{eq:scalin}): 
\begin{equation}
\mathcal F_{\mathrm{lin}}(X) \sim  \frac{X^{\tau_q}}{[V_g(X)]^{q-1}}, \; \text{for }X\to \infty, W<W_c \; .
\end{equation}
 
In the other volumic scaling hypothesis \eqref{eq:scavol}, the scaling parameter $\Lambda(W)$ plays the role of the correlation volume, and the scaling function is expressed as the ratio of the two characteristic volumes instead of the ratio of the characteristic linear sizes. In this case, a critical structure is an hypercube, and the number of such structures in a delocalized state is $N/\Lambda$. Therefore, the asymptotic behavior of $P_q$ when $N\gg \Lambda$ is 
\begin{equation}
\langle P_q \rangle \approx \left(\frac{\Lambda}{N}\right)^{q-1} \left(1-\tau_q \frac{\xi}{d_N}\right)\; .
\end{equation}
This is precisely what is predicted by the analytical theory \cite{S-mirlin1994distribution}: $P_2 \approx C/N$ with $\ln C\propto (W_c-W)^{-\kappa}$ close to the threshold, and $\kappa$ the correlation length critical exponent. The corrections in the parenthesis are negligible in the limit $d_N \gg \xi$. In this case, the scaling function $ \mathcal F_{\mathrm{vol}}$ has the following asymptotic behavior: 
\begin{equation}
\mathcal F_{\mathrm{vol}}(X) \sim  \frac{\left[{V_g}^{-1}(X)\right]^{\tau_q}}{X^{q-1}} , \; \text{for }X\to \infty, W<W_c \; .
\end{equation}

Finally, the volumic scaling hypothesis for the NEV $\lambda(N,W)$ 
\begin{equation}
\frac{\nev}{N/d_N} = \mathcal G_\mathrm{vol} \left( \frac{N}{\Lambda} \right) \; ,
\end{equation}
with $d_N\sim \log_2 N$, implies the following asymptotic behavior:
\begin{equation}
\mathcal G_{\mathrm{vol}}(X) \sim  \frac{\left[{V_g}^{-1}(X)\right]}{X} , \; \text{for }X\to \infty, W<W_c \; .
\end{equation}
Because ${V_g}^{-1}(N/\Lambda)=d_N - \xi$, in the limit of large system volume $N\gg \Lambda$:
\begin{equation}
 \lambda \approx \frac{N}{d_N} \frac{\Lambda}{N} (d_N-\xi) = \Lambda \left(1-\frac{\xi}{d_N}\right) \; .
\end{equation}
Therefore, in the delocalized phase, the NEV $\lambda$ saturates to the correlation volume in the thermodynamic limit.

\subsubsection{Correspondence between the behaviors of the moments versus $N$ or versus $\ell$}

\begin{figure}
  \includegraphics[width=0.4\textwidth]{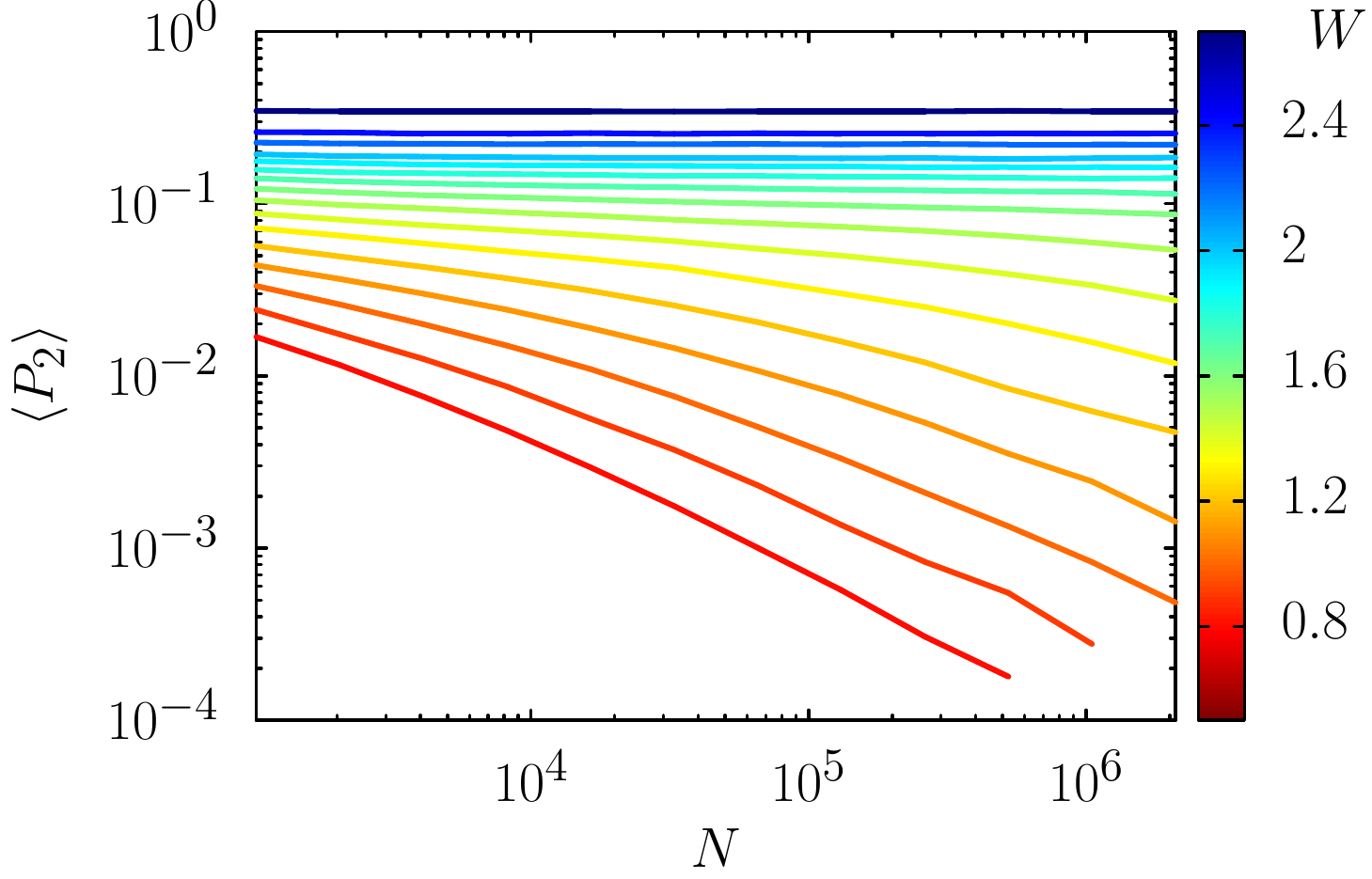}
  \caption{Average moments $\langle P_2 \rangle$ as a function of the volume $N$ for $p=0.06$ and $W\in [0.8,3]$. The data correspond to Fig.~\ref{fig:scalingP2}}
  \label{fig:rawSupMat}
\end{figure}

The figure \ref{fig:rawSupMat} shows the moments $\langle P_2 \rangle$ without any coarse graining ($\ell=1$) as a function of $N$ for different $W$. Three regimes are observed: for $W>W_c\approx 1.6$ and $d_N\sim \log_2 N\gg \xi$, a localized behavior corresponds to $\langle P_2 \rangle$ independent of $N$. 
For $W \approx W_c$ and $N \ll \Lambda$, the critical strongly non-ergodic regime manifests itself as a quasi-plateau where $\langle P_2 \rangle \sim \log_2 N^{-\tau_2}$, whereas for $W<W_c$ and $N>\Lambda$ the ergodic regime manifests itself as an asymptotic decrease of $\langle P_2 \rangle$ as $1/N$.

 In the box-counting method used in Fig.~\ref{fig:P2deel}, we observe similar behaviors: For $W <W_c$, at large scales ($l \gg \Lambda$), an ergodic behavior manifests itself as $\langle P_2 \rangle \sim \ell$. At scales $\ell \ll \Lambda$, the critical strongly non-ergodic behavior corresponds to a quasi-plateau in $\ell$. However, the multifractal regime at scales below $1/(2p)$ of Fig.~\ref{fig:P2deel} cannot be seen in Fig.~\ref{fig:rawSupMat}, since $N< 1/(2p)$ would correspond to a system with no long range branching.

\end {document}